# Urban transfer entropy across scales

**Roberto Murcio[1*], Robin Morphet[1], Carlos Gershenson[2], Michael Batty[1]**

[1] Centre for Advanced Spatial Analysis, University College London, London, UK

[2] Instituto de Investigaciones en Matemáticas Aplicadas y en Sistemas, Universidad Nacional Autónoma de México, Distrito Federal, México

The morphology of urban agglomeration is studied here in the context of information exchange between different spatio-temporal scales. Urban migration to and from cities is characterised as non-random and following non-random pathways. Cities are multidimensional non-linear phenomena, so understanding the relationships and connectivity between scales is important in determining how the interplay of local/regional urban policies may affect the distribution of urban settlements. In order to quantify these relationships, we follow an information theoretic approach using the concept of Transfer Entropy. Our analysis is based on a stochastic urban fractal model, which mimics urban growing settlements and migration waves. The results indicate how different policies could affect urban morphology in terms of the information generated across geographical scales.

*r.murcio@ucl.ac.uk
r.morphet@ucl.ac.uk
cgg@unam.mx
m.batty@ucl.ac.uk

# Introduction

In this paper we examine the transfer of information between two spatial levels in a spatial hierarchy over time. The two spatial levels are termed regional and local reflecting their level of resolution. The change model that we use to show change over time represents the temporal dynamics and has three main components, a percolation process, a diffusion process and a criticality process. The percolation model describes the changing character of the cells as they become occupied. The diffusion takes place over the cells, one of which acts as a seed. As migrants diffuse the cell occupancy increases up to a threshold after which the sandpile, that is the accumulation of growth, collapses thus spreading the capacity of the cell in question to its adjoining cells. At the termination of the process we have migrants occupying many of the cells. The analysis then proceeds by identifying clusters using a subtractive clustering algorithm. With clusters defined at regional and local levels we are then in a position to calculate the transfer entropy between the two levels in both directions, regional to local and local to regional, thus providing us with a new measure of spatial dependence.

Conventional models of metropolitan areas are usually confined to an analysis at one scale or level [1, 2] although interactions and migrations between the scales may be of importance in understanding the dynamics of the modelled system [3]. Classical theory [4] defines three conditions that must be met for migration to occur:

(a) *Complementarity:* there must be a benefit in locating in the desired destination relative to the origin

(b) *Intervening opportunity:* where there is potential movement from an origin to a destination intermediate competing destinations must also be taken into account.

(c) *Migration cost:* this factor could be seen as a distance/friction parameter. If the cost of moving between an origin and a potential destination is too great, movement will not take place, regardless of conditions (a) and (b).

We will embody these principles in the hypothetical urban model that we use to explore interaction between spatial and temporal scales measuring these interactions using various information measures that we will now describe.

# Information Measures

We define below the measure that we use for analysis of the clusters, namely transfer entropy. In information theory [5], the Shannon entropy represents the basic measure of information and this is defined as,

$$H(X) = -\sum_{x \in X} p(x) \log p(x) \quad , \quad \sum_{x} p(x) = 1 \tag{1}$$

which is the preferred measure for detecting the reduction in uncertainty by any measurement $x$ of a random variable whose probability is $p(x)$. Extending Shannon entropy to measure the uncertainty between two interacting random variables $X$ and $Y$, at different temporal or spatial scales for example, is accomplished using mutual information, $I(X,Y)$, defined by

$$I(X,Y) = \sum_{x \in X} \sum_{y \in Y} p(x,y) \log \frac{p(x,y)}{p(x) \cdot p(y)} \tag{2}$$

One drawback with $I(X,Y)$, is its lack of directionality as, $I(X, Y) = I(Y, X)$, which, in our analysis, would imply that the future has a causal effect on the past. Mutual Information can also be expressed as the difference between two entropies thus

$$I(X,Y) = -\sum_{x \in X, y \in Y} p(x,y) \log(p(x) \cdot p(y)) + \sum_{x \in X, y \in Y} p(x,y) \log(p(x,y)) \tag{3}$$

with the first term of the right hand side representing the entropy assuming independence and the second representing the observed entropy.

Transfer entropy, *TE*, was developed by Schreiber [6] to overcome the time symmetric limitation of mutual information. Given two sample spaces of information, $X = \{x_1, x_2, ..., x_t\}$ and $Y = \{y_1, y_2, ..., y_t\}$, the transfer entropy from $X$ to $Y$, is obtained from defining the entropy rate between two systems as the amount of additional information gained from the next observation of one of the two systems.

Following [6], let us consider two systems, $X$ and $Y$ and define two entropies. In the first case we define an entropy (actually an entropy rate as it depends on time $t$, based on the assumption that $y_{t+1}$ depends on both $x_t$ and $y_t$

$$h1 = -\sum_t p(y_{t+1}, y_t, x_t) \log(p(y_{t+1} | y_t, x_t)) \tag{4}$$

In the second case we define an entropy in which $y_{t+1}$ depends only on $y_t$ thus

$$h2 = -\sum_t p(y_{t+1}, y_t, x_t) \log(p(y_{t+1} | y_t)) \tag{5}$$

Equation(4) defines the entropy given dependence on $x_t$ and $y_t$ whilst equation(5) shows dependence on $y_t$ only.

Transfer entropy, $T_{XY}$, the transfer of information from X to Y, is then defined as the difference between these two rates in the same way that mutual information was defined in equation(3). Then

$$\begin{aligned} T(X,Y) &= h2 - h1 \\ &= \sum_{t=1} p(y_{t+1}, y_t, x_t) \log\left(\frac{p(y_{t+1} | y_t, x_t)}{p(y_{t+1} | y_t)}\right) \\ &= \sum_{t=1} p(y_{t+1}, y_t, x_t) \log\left(\frac{p(y_{t+1}, y_t, x_t) \cdot p(y_t)}{p(y_t, x_t) \cdot p(y_{t+1}, y_t)}\right) \end{aligned} \tag{6}$$

This measure of information transfer resembles the Kullback-Leibler distance but applied to conditional probabilities. The transfer entropy $T(Y,X)$ can be derived in a similar fashion giving

$$T(Y,X) = \sum_{t=1} p(x_{t+1}, x_t, y_t) \log\left(\frac{p(x_{t+1}, x_t, y_t) \cdot p(x_t)}{p(x_t, y_t) \cdot p(x_{t+1}, x_t)}\right) \tag{7}$$

The asymmetry is confirmed as equation (6) does not equal equation (7).

## The Urban Model

Our application is to the United Kingdom where we plant the first seed of growth at the historic location of the City of London [7], one of the earliest major settlements in the country, the first encampment of the Roman army in its invasion of Britain in 53AD. This represents the initial historical event which we use to initiate our modelling of the settlement of the UK. We model urban growth using two well-known fractal processes: percolation and diffusion limited aggregation which is linked to the self-organising process defined by Bak,

Tang and Wiesenfeld (BTW) [8]. The BTW process converges to an attractor, the critical point of which is reached from a wide variety of starting conditions. At the critical point the process said to have achieved self- organised criticality (SOC). Our urban model aims to capture the different spatial patterns and dynamics observed in a system of settlements at different geographical scales. For regional scales (~1:1,500,000) we applied diffusion and percolation in order to represent two of the main drivers of urban growth: migration of the population and the economics of agglomeration respectively. The morphology at local scales (~1:200,000) derives from imposing SOC characteristics on the model.

A regular lattice was set up to cover the study area with a grid. This grid represents the available land for urban growth and occupies some 302x388 cells. Each of these cells, $i$ represents a region of the physical terrain of approximately 4.8 km$^2$. Our simulations run through a series of discrete steps, represented by time $t$, and at each step, $U_i(t)$, an urbanisation index for each cell is defined by: non-urbanized ($U_i(t)=0$; urbanised consolidated ($U_i(t)=1$); and urbanized non-consolidated ($U_i(t)=2$). The meaning of these values and their derivation is described below. They are used in the analysis to distinguish different types of cluster.

## The percolation process

Vicsek and Szalay proposed a cellular automaton model to study the fractal distribution of galaxies [9]. In their lattice model, a cell *i* which represents a mass element, would become part of a galaxy based on two parameters; the potential of belonging to a galaxy at position *i*, and a given threshold that regulates such a potential. Fujita and Thisse in their study of economic agglomeration theory [10] say that *"... just as matter in the solar system is concentrated in a small number of bodies (the planets and their satellites); economic life is concentrated in a fairly limited number of human settlements (cities and clusters). Furthermore, paralleling large and small planets, there are large and small settlements with very different combinations of firms and households."* Following these ideas, a percolation model to study the fractal distribution of galaxies based on [11] is applied here to model the development of urban settlements.

We consider a potential for urbanization $P_i(t)$ in cell *i*. As development takes place more cells develop the potential to urbanize and the decision on whether cell *i* becomes urbanized or not depends on an external parameter, the development threshold *T*. If $P_i(t) > T$ and cell *i* has not already undergone urbanization, then cell *i* becomes urbanized and is consolidated with $U_i(t) = 1$. Once a migrant unit is attached to an urban settlement at *i*, it could be the case that for this particular *i*, $P_i(t)$ is not greater than *T*; if that is the case, the cell *i*, occupied by the migrant unit would became an urbanized non-consolidated urban settlement with $U_i(t) = 2$; this reflects a pattern of development in many of the urban belts formed around metropolitan areas. If after recalculation of $P_i(t)$ a migrant unit arrives at a cell with $U_i(t) = 2$ and $P_i(t) > T$, then the value of $U_i(t)$ becomes equal to 1. The potential $P_i(t+1)$ is defined as:

$$P_i(t+1) = \sum_{j \in \Omega} \frac{P_j(t)}{5} + K.C_i(t) + \varepsilon_i(t) \qquad (8)$$

where $\Omega$ represents the cell, *i*, plus its von Neumann neighbourhood[1] and $\varepsilon_i(t)$ is either 1 or −1 with equal probability. The inclusion of random events prevents any tendency towards equilibrium since noise is continually being introduced into the system. Once cell *i* becomes urbanized and consolidated, it cannot reverse this process, even if its potential $P_i(t)$ falls below the value of *T*.

The function $C_i(t)$ calculates the current number of urban units (buildings for example) that a cell *i* has at a particular time *t*. The constant $K > 0$ which is a damping factor, is fixed at the beginning of the simulation and represents the influence that the local settlements have over the regional potential (*K=0.1* in our simulations). The formulation for $C_i(t)$ is given in equation(11).

## The diffusion process

As discussed in the introduction, the conditions for migration to occur are fulfilled by a diffusion limited aggregation-like process [12]. At each time *t* (Eq. 11), we select a fixed number of random positions over the study area and at each of these positions we create a migrant unit that begins a walk (diffusion) over any configuration that is emerging at that particular *t* until reaches some cell *i*, being in the nearest neighbourhood of an already occupied cell *j*. The nature of the walk is address in section 3.3. The number of migrant units $N(t)$ created is not fixed in all the simulations but is defined from:

$$N(t) = \begin{cases} 0 \text{ for } t < 50 \\ t \bmod 10 \text{ for } 50 \leq t \leq 500 \end{cases} \tag{9}$$

so $N(t)$ increases over time.

---
[1] The north, south, east and west cells of the central cell *i*.

*3.3 Self-Organized Criticality*

Self-Organized Criticality (SOC) is a characteristic possessed by spatially extended dynamical systems and other complex systems [13, 14]. When a phenomenon changes state, the resulting reactions are distributed across time and space at all levels in a way that can range from a simple isolated movement, to chain reactions involving all activities in the system. A good example is the sand pile model, as presented in [14] and, based on this we introduce SOC into our model. However, instead of sand grains, we have urban units that migrate and in doing so, may trigger a chain reaction in which other units also relocate. To model this, two parameters are required for each cell *i*, a maximum capacity $C_i^{\max}$ and a current capacity $C_i(t)$. The $C_i^{\max}$ parameter represents the maximum number of urban units that cell *i* can hold and is defined as

$$C_i^{\max}(r) = C_{seed}^{\max} \cdot r^{-\alpha} \tag{10}$$

where *r* is the Euclidean distance from the spatial position of the seed to cell *i* and, $C_{seed}^{\max}$ is the maximum capacity for the historical accident, which is set here to 100 to ensure that its capacity is never constrained, and $\alpha$ is a parameter of the distance distribution which we define as a power law. Equation (10) constrains the density of an area by how far it is from the seed. The exponent $\alpha$ (which is set to 0.3 in our simulations) can be thought as a density gradient held constant over time. In reality appears to decrease gradually as the city grows [11]. In terms of the sand pile model, the critical slope of the distribution of the urban units over an area is controlled by $\alpha$.

Once we have calculated the $C_i^{max}$ for every cell $i$, (except the seed cell which has fixed capacity) we need to set up the current capacities, $C_i(t)$, for the whole lattice excepting the seed cell which is equal to 1, as we assume that at least one urban unit is already there.

Capacities are then defined as follows in equation (11) which represents the sand pile process. If $C_i(t) > C_i^{max}$ for a cell $i$, then an avalanche begins (as in the sand pile model) around cell $i$ by its losing 4 urban units to its von Neumann neighbourhood. The damping constant now defined as $\kappa > 0$, similar to that introduced in equation(8), is fixed at the beginning of the simulation with a value of *0.1* and represents the influence that the regional settlements have over the local capacities. The value of K and $\kappa$ ensures that the analysis of asymmetric regional and local influence is not biased by a varying damping factor. Cell capacities are updated using the following algorithm

$$C_i(t+1) = \begin{cases} 0 \text{ for } t=0 \\ C_i(t)+1 \text{ for } t \neq 0 \text{ and } P_i(t+1) > P_i(t) \\ C_i(t)-4 \text{ and } C_\Omega(t+1) = C_\Omega(t)+1+\kappa P_i(t) \text{ for } \begin{cases} t \neq 0 \\ \text{and} \\ C_i(t) > C_i^{max} \end{cases} \end{cases} \quad (11)$$

# Running the model

We defined a set of four thresholds *T* equal to *4.0, 4.5, 5.0* and *6.0* relating to the difficulty of development actually occurring, which are determined by physical or policy factors. We fixed the length of the simulation to *t=500* iterations as this gives sufficient time for patterns to emerge. At time *t = 0* all grid positions, except the seed, have a potential $P_i(0) = \varepsilon_i$. To represent the importance of the historical accident across time, its initial potential is $P_{seed}(0) = 20$ which remains constant through all the iterations, so equation(8) is never applied to the seed cell and its potential always exceeds the threshold. A typical configuration obtained with this model is shown in Fig. 1. As our approach is stochastic we performed 1000 runs per configuration in order to derive robust statistics. All the quantities and measures derived are then the averaged over each configuration. We refer the reader to [15, 16] for examples in which aspects of this model have been applied.

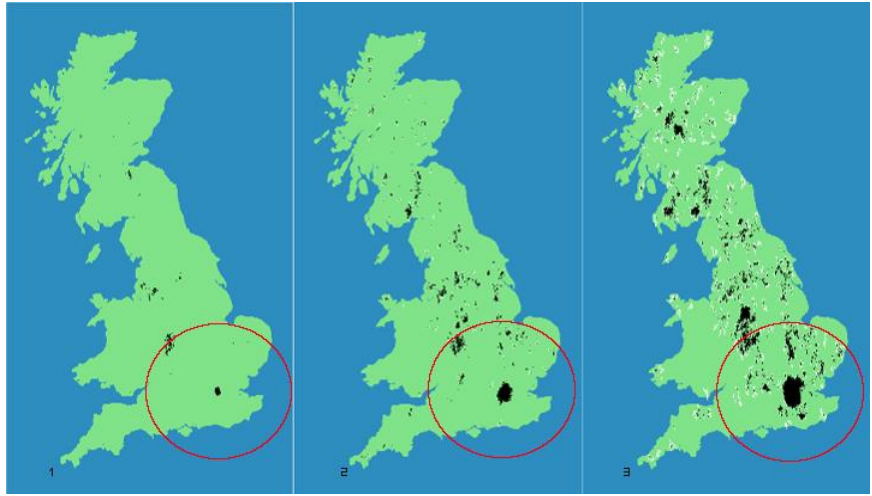

Fig. 1. Three different stages in our model evolution. From left to right we can observe structures formed at (1) $t_1$ *(t=50)* (2) $t_5$ *(t=250)* and (3) $t_{10}$ *(t=500)*. Consolidated structures $C_i(t) = 1$ are shown in black, while non-consolidated ones $C_i(t) = 2$ are shown in white. The red circumference around the London area is the area of influence defined around the initial seed.

## Urban migration

Migration units walk across the lattice (as described in section 3.1) until they fix their position in a cell *i* according to the rules set out above. This walk is achieved by selecting a subset of all possible trajectories that migrant units may take. We accomplish this in two different ways:

a) All migration units scan a defined area in search of the location with the highest potential, and when they find it, they walk a certain distance in terms of their units towards that location;
b) An area of influence with a radius of 100 cell units, centred at the seed is created. Once a migration unite enters this zone, it will prefer to move towards the seed.

# Analysis

## Time Series Construction

A ten point time series is defined for the configurations obtained by our model. This was accomplished by recording the pattern of urban structures generated at times $t = \{50, 100, 150... 500\}$. We label each point in the time series as $t_j$, with $t_1=50$, $t_2=100$, until $t_{10}=500$. The main problem here is how to assign them a unique value in order to apply Equation (4). We performed a classification analysis based on the number of clusters **N** detected at each $t_j$, applying a subtractive cluster algorithm [17] that strongly depends on a predefined cluster radius $r_c$ which is used as a cut-off to define these time series at different geographical scales. The logic is as follows: at a local level, we can observe in great detail the urban structures that surround us and easily distinguish one structure from another, but our observation area is very limited (based on the small radius of $r_c$); as we begin to increase the geographical scale, much of the urban detail begins to disappear, as many structures become indistinguishable from each other but now our observation area covers many more structures (based on a larger radius $r_c$).

The subtractive cluster algorithm can be summarized in three steps:

1. Select the data point $k$ with the highest potential $G$ to became a cluster or group centre, according to

$$G_k = \sum_{j \in k-j \leq r_c} e^{-\beta \|k-j\|^2} \qquad (12)$$

2. Where $\beta = \dfrac{4}{r_a^2}$ and $r_a$ is a positive constant and $\|k-j\|$ is the Euclidean distance between points $k$ and $j$.
3. All data points in point $k$'s $r_a$ vicinity are labelled as one cluster and removed from further calculations.
4. The process continues iterating on steps 1 and 2 until all the data is within radii $r_a$ of a cluster centre

In our calculations, parameter $r_a$ takes values in the range {1, 46.6, 92.3, 138, 183.6, 229.3, 275, 320.6, 366.36, 412}. This selection is constrained by the maximum (1) and minimum (412) Euclidean distance that a pair of cells *i* and *j* can have in our lattice. The intermediate values are calculated dividing 412 by 10, as we are defining 10 different geographical scales. For example, for *T=4.5* we obtained the time series shown in Table 1.

**Table 1.** Times series for *T=4.5* showing the number of clusters generated at each scale

| Scale | Time periods | | | | | | | | | |
|---|---|---|---|---|---|---|---|---|---|---|
| | $t_1$ | $t_2$ | $t_3$ | $t_4$ | $t_5$ | $t_6$ | $t_7$ | $t_8$ | $t_9$ | $t_{10}$ |
| 1 | 51 | 87 | 154 | 230 | 315 | 408 | 502 | 607 | 732 | 873 |
| 2 | 4 | 5 | 7 | 9 | 11 | 14 | 16 | 20 | 24 | 29 |
| 3 | 3 | 4 | 6 | 7 | 8 | 9 | 10 | 11 | 12 | 13 |
| 4 | 3 | 4 | 5 | 6 | 7 | 7 | 7 | 8 | 8 | 9 |
| 5 | 3 | 4 | 5 | 5 | 4 | 6 | 6 | 6 | 7 | 7 |
| 6 | 3 | 3 | 4 | 4 | 4 | 4 | 5 | 5 | 5 | 5 |
| 7 | 3 | 3 | 3 | 4 | 4 | 4 | 4 | 4 | 4 | 4 |
| 8 | 2 | 3 | 3 | 3 | 3 | 3 | 3 | 3 | 3 | 4 |
| 9 | 2 | 3 | 3 | 3 | 3 | 3 | 3 | 3 | 3 | 3 |
| 10 | 2 | 2 | 2 | 2 | 2 | 2 | 2 | 2 | 2 | 2 |

As the scale increases, the number of cluster between time periods became more and more stable, because the urban structures became undistinguishable from each other at larger scales. At scale 10, we have only two big clusters representing the whole study area at any time period.

Using the previous method, as well as the results shown in Table 1, another two tables were constructed giving a total of four tables, one for each T.

## Transfer Entropy calculations

In equation (6) the joint probability calculation is key to obtaining the *TE* value. Estimating these probabilities has been proved to be a very difficult task. Several methods via probability density function estimation have been proposed to solve this problem [19]. In this research, representing the different scenarios, through the six tables mentioned above, we calculated the empirical joint probabilities for the emerging configurations, i.e., we calculated the probability of having a certain number of clusters between scales, assuming that if a particular configuration is not found in a particular table, and then the joint probability for that configuration is zero. Thus, if we take Table 1, in order to calculate the $TE_{Scale\ 1 \rightarrow Scale\ 2}$ at step $t_1$, we need to find out the probability $p(y2,y1,x1)$ as required by the first term of equation (8). This probability is easily obtained inspecting the first and second rows of the mentioned table (as Scale 1 is represented by the first row and Scale 2 by the second row). The term $y_2$ corresponds to the value located in the second column, second row; $y1$ is the value at the first column, second row and $x1$ is located in the first column, first row position, i.e., $p(y2,y1,x1) = (5,4,51)$. We then count how many combinations of these values exist in these two rows. For this example there is only one, so $p(y2,y1,x1) = (5,4,51)=1/10$, as there are 10 positions in the time series. Now, if we take $TE_{Scale\ 7 \rightarrow Scale\ 8}$ and $t_4$, then we examine rows 7 and 8 to find the values for *y5,y4* and *x4*, which are 3, 3 and 4, so, $p(y4,y3,x3) = (3,3,4)=5/10$, as there are seven identical (3,3,4) combinations between rows 7 and 8.

## Results and Discussion

Taking Table 1 again as a typical cluster configuration, we observe that at time $t_1$, except for Scale 1 (the most local scale), the number of clusters at all scales is very low, because at this point almost no urban structures are formed; for at time $t_{10}$ the number of clusters reported is constant, which suggests that when the urban structures reach the point of consolidation, the probability of detecting new developments is low; on the other hand, the change between $t_1$ and the rest of the temporal sequence is dramatic, implying at this early stage, the full complexity of the urban structures is only measurable at a local scale.

To construct a coherent analysis for the *TE* results among all our scales and thresholds, they were categorised as:
     a) *TE* between contiguous scales (scale 1 vs scale 2, scale 2 vs scale 3, etc.)
     b) *TE* between non-contiguous scales (scale 1 vs scale 3, scale 1 vs scale 4, etc.).
This shows how much of the information generated at one scale *i* is responsible for the information obtained at scale *j* and how the middle scales filter such information. As in the analysis of clusters above, the *TE* values obtained reflect the differences between different thresholds. This suggests that as settlements become harder to generate, as a result of policy or physical factors, the probability that a migrant unit settles, decreases since it becomes harder for it to find a consolidated area to attach to. In reality, these migrant units would still settle somewhere, regardless of the urban policy, but in our model this fine grain detail is not taken into account.

In general, the *TE* in the local→regional direction dominates that for the regional→local, as is show in Fig. 2. Information flows less easily from regional to local scales, a situation that might be expected as the information generated at the higher scale of the urban system is not ultimately responsible for the creation of the local structures; policies established at a regional level, unless they are very restrictive and impose a particular local development pattern, would tend to become diluted (and be less effective) at the more local scales. One notable aspect of all the plots in Fig. 2 is that the *TE* rises initially with the length of the rise decreasing as the threshold increases. At the early stages of urban structure formation, the system always increases the transfer of information in both directions. But this tendency does not hold as the system continues to grow. As the urban systems get larger, the information transfer tends to equalise with, in the higher threshold cases, the regional→local direction dominating that for the local→regional as the scale increases. This equalisation may simply represent the convergence of the two $r_a$ values at each mark as the mark number increases but it suggests that the information transfers at the higher scales are less significant in both their absolute size and in their directional difference. Where the dominance reverses, the difference in information flows is small suggesting a similarity in the cluster pattern. This might be expected if growth was concentrated mainly at the cluster edges and the increase in $r_a$ left the number of clusters substantially unchanged. The peaks in *TE* may therefore be seen as the points at which the growth pattern changes from one of many new separated clusters to fewer larger clusters caused initially by edge accretion of cells and in the later stages by amalgamation of larger clusters. This would account for the multiple peaks observed in the local→regional *TE*. Interestingly enough, for *T=4.5*, mark 4, *TE* values, in both directions equal zero, meaning that the number of clusters detected from scale 4 to scale 5 is exactly the same. There was no particular change in the overall structure of the system from one scale to the other. Further research is needed concerning this effect.

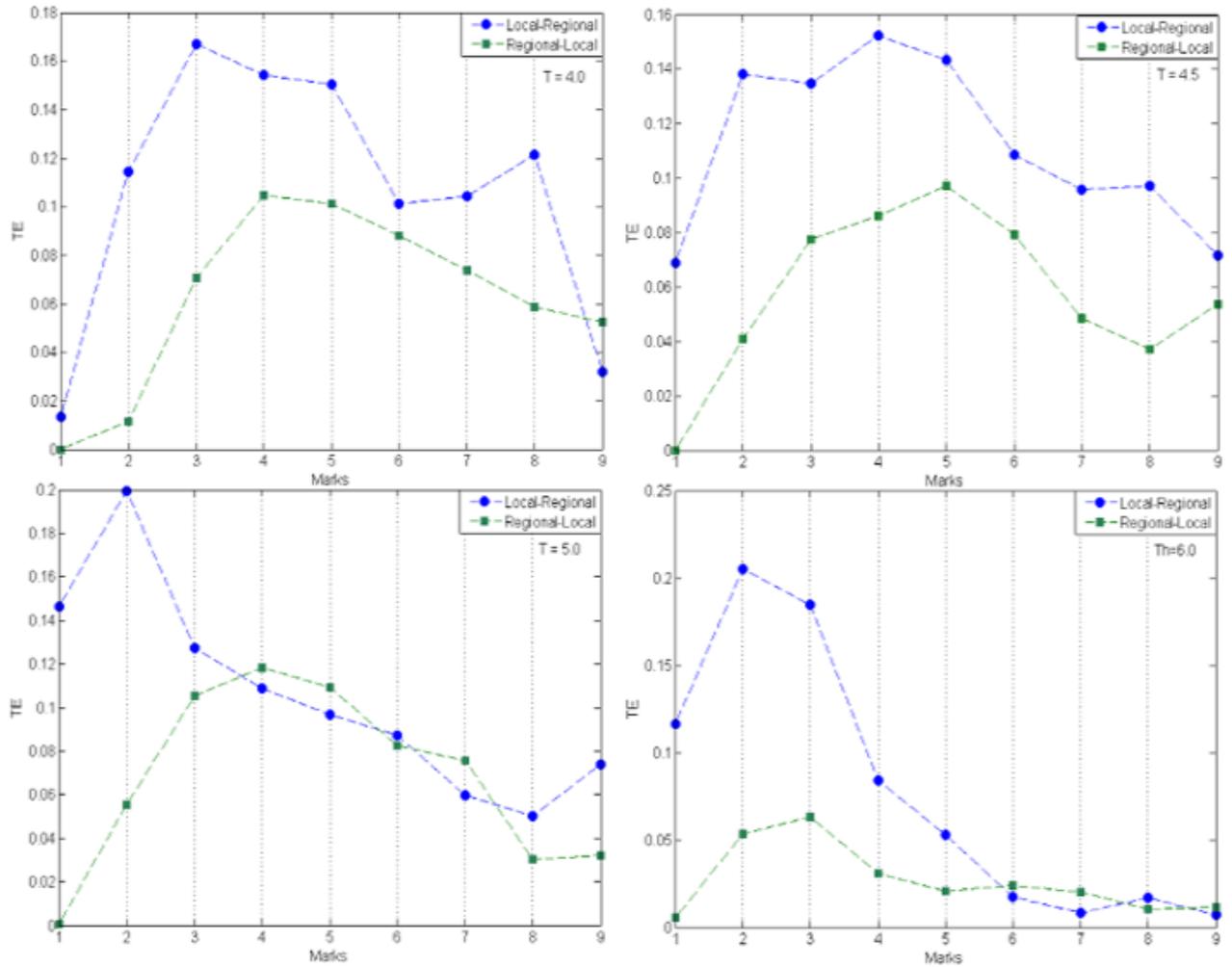

*Fig. 2. TE* between contiguous scales. The x axis should be read as follows: Mark 1–*TE* Local→Regional (Scale 1, Scale 2) and *TE* Regional→Local (Scale 2, Scale 1); Mark 2–*TE* Local→Regional (Scale 2, Scale 3) and *TE* Regional→Local (Scale 3, Scale 2), etc.

For the threshold *T=5.0*, representing a moderate urban policy at mark 4, *TE* values for the regional→local direction are greater than the local→regional until mark 8 (except for mark 6 in which both values are practically equal). At these points, the structure of the information changes in a way that allows the information generated at higher scales to flow to lower ones. From mark 6 this is not that surprising, because we are operating at the higher scales and the information generated at a regional level is mainly responsible for the structures observed at this level. At mark 4 (*TE* between scales 4 and 5, in both directions) the net direction of the information is indistinguishable in practical terms. This suggests that the regional pattern of information is leading the local but this may simply be the effect of scale with the regional scale identifying as clusters those settlements at a local scale which are growing by accretion. In Fig. 3 we show a section from one typical configuration generated for scales 4 and 5,

*T=5.0*. Although the latter contains a greater proliferation of urban structure, it does not seem to contain more clusters, thereby reinforcing the view that the *TE* equality reflects growth by accretion and amalgamation with little change in the cluster numbers. The reversal of dominance and the near equality of *TE*s from thresholds 4.0, 4.5 and 5.0 reflect a change in the distribution of the settlements.

Finally, for *T=6.0*, this represents a more restrictive urban policy. Once the scales begin to increase, the number of clusters generated at this threshold begin to stabilize, so when we calculate the *TE* between series it tends to zero. As stated above, when $T \to \infty$ the number of clusters tends to zero so the *TE* would also tend to zero in the process. This unrealistic scenario implies restrictive urban policies, to the point where no urban structures could be created. Alternatively, when $T \to 0$, (no restrictions at all) this would lead to same result since each location can be urbanized, giving one big cluster with zero *TE*.

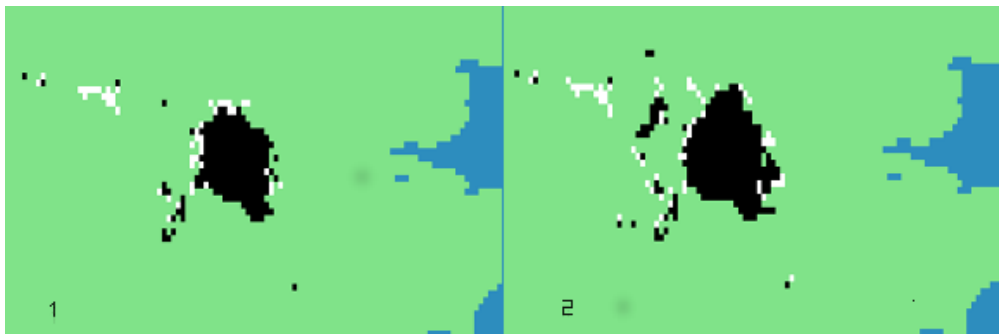

*Fig. 3*. The London area zoom generated with our urban model for *T=5.0* . (A) $t_4$ (B) $t_5$

The last *TE* measure performed was over the non-contiguous scales. Fig. 4 presents 10 plots at *T=5.0*. First, notice that from Scale 2, all plots, in both directions, have a zero value. This point represent the *TE* (Scale i, Scale i), so it should not be considered as part of the analysis *per se*. We kept this in order to generate a continuous sub-plot.

These plots show us in one single frame the behaviour between directions of information flow. From sub-plot Scale 3 onwards, the regional→local direction *TE* values dominate where they are to the left of *TE*=0. Conversely to the right of *TE*=0, the local→regional tend to dominate. The division where we lose the non-consolidated structures from the analysis is very clear: until Scale 5, there is a constant gap at the right part of the *TE* zero value, while from Scale 6, this gap can be found on the left part. The fact that the *TE* value between directions is kept constant is evidence that the amount of information that is flowing from one scale to the rest does not get amplified or decreased at these bands. From Scale 6, the situation is similar, but mirroring at the left part of the *TE* zero value and at the right part, the values between directions are very close to each other, as we have already established at this scale.

The other somewhat unusual result is that the *TE* between Scale 1 every other, in the regional→local direction, is practically zero. All the information generated at the regional scale never reaches the local level, a situation that is also reflected in all the other thresholds. This situation is observed on a daily basis in our cities, where political decisions, imposed at the regional level, never reach or find their way down to improve the lives of citizens at local level. The current debate in British politics and government for example, is largely about these issues of regionalism and localism.

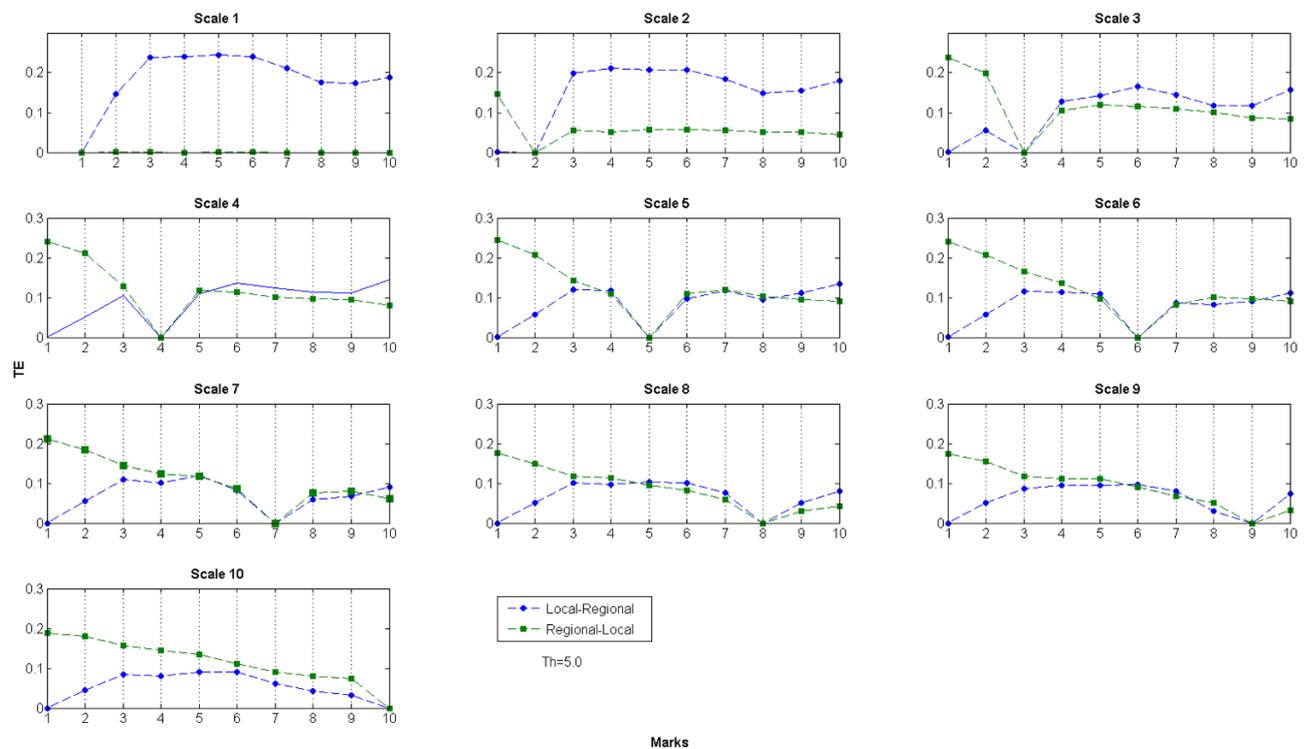

Fig. 4. TE between non-contiguous scales, *T*=5.0. Each one of the 10 sub-plots presented is labelled as Scale i (i from 1 to 10), meaning that this particular plot represents the values for the *TE* between Scale i and the rest of the scales. The marks axis should be read as follows: 1–*TE* Local→Regional (Scale i, Scale 2) and *TE* Regional→Local (Scale 2, Scale i); 2–*TE* Local→Regional (Scale i, Scale 3) and so on until 10–*TE* Local→Regional (Scale i, Scale 10) and *TE* Regional→Local (Scale 10, Scale i).

# Conclusions

The information transfer between scales *i* and *i+n*, cannot be constructed as the sum of the transfer information between scales *i* and scale *i+1* plus scale *i+1* and scale *i+2*...scale *i+j* and scale *i+n,* with *1<j<n*. This is one of the classic footprint for a complex system. It shows that urban patterns are not reducible to description at a single scale, but require a multi-scale description. This asymmetry reflects the hierarchical structure of spatial analysis where questions of scale are important. The model identifies London, the West Midlands, and the North West as urban centres, which is reassuring given the simplicity of the assumptions. Our results support the idea that decisions or information are flowing more easily from the local scale to the regional than the other way around. This is because *TE* is higher from lower scales to higher scales than vice versa. Transfer entropy seems a promising analytic tool for examining the effect of scale but requires further testing against a range of assumptions for *K* and *κ* and for the area of influence. Alternatively, these could be defined endogenously.

The loss of the non-consolidated structures from the calculations appears as the key factor in this change of regime or phase between contiguous scales and may prove policy relevant particularly in those studies where competition between regulated and unregulated development is of importance.

There are still outstanding questions which we need to explore further. Further work would include comparisons with real data, studying lower scales [20], other urban regions, and more sophisticated models to test different hypotheses about the effects of local vs. regional urban planning and growth. A similar asymmetric flow of information is also likely to be seen in organisational and social hierarchies and this offers some scope for an integration of social and spatial analysis.